\documentclass{article}

\usepackage{microtype}
\usepackage{graphicx}
\usepackage{subfig}
\usepackage{amsmath}
\usepackage{amssymb}
\usepackage{booktabs} 
\usepackage{url}
\usepackage{graphicx}
\usepackage{hyperref}

\newcommand{\theHalgorithm}{\arabic{algorithm}}

\usepackage[accepted]{icml2021}

\icmltitlerunning{VARA-TTS: Non-Autoregressive Text-to-Speech Synthesis based on Very Deep VAE with Residual Attention}

\begin{document}

\twocolumn[
\icmltitle{VARA-TTS: Non-Autoregressive Text-to-Speech Synthesis based on \\ Very Deep VAE with Residual Attention}



\icmlsetsymbol{equal}{*}

\begin{icmlauthorlist}
\icmlauthor{Peng Liu}{tencent}
\icmlauthor{Yuewen Cao}{equal,cuhk}
\icmlauthor{Songxiang Liu}{equal,cuhk}
\icmlauthor{Na Hu}{tencent}
\icmlauthor{Guangzhi Li}{tencent}
\icmlauthor{Chao Weng}{tencent}
\icmlauthor{Dan Su}{tencent}
\end{icmlauthorlist}

\icmlaffiliation{tencent}{Tencent AI Lab}
\icmlaffiliation{cuhk}{Human-Computer Communications Laboratory, The Chinese University of Hong Kong}

\icmlcorrespondingauthor{Peng Liu}{laupeng1989@gmail.com}

\icmlkeywords{Machine Learning, ICML}

\vskip 0.3in
]



\printAffiliationsAndNotice{\icmlEqualContribution} 

\begin{abstract}
This paper proposes VARA-TTS\footnote{Codes will be released soon.}, a non-autoregressive (non-AR) end-to-end text-to-speech (TTS) model using a very deep \textbf{V}ariational \textbf{A}utoencoder (VDVAE) with \textbf{R}esidual \textbf{A}ttention mechanism, which refines the textual-to-acoustic alignment layer-wisely. 
Hierarchical latent variables with different temporal resolutions from the VDVAE are used as queries for the residual attention module.
By leveraging the coarse global alignment from previous attention layer as an extra input, the following attention layer can produce a refined version of alignment. This amortizes the burden of learning the textual-to-acoustic alignment among multiple attention layers and outperforms the use of only a single attention layer in robustness. 
 An utterance-level speaking speed factor is computed by a jointly-trained speaking speed predictor, which takes the mean-pooled latent variables of the coarsest layer as input, to determine number of acoustic frames at inference.
Experimental results show that VARA-TTS achieves slightly inferior speech quality to an AR counterpart Tacotron 2 but an order-of-magnitude speed-up at inference; and outperforms an analogous non-AR model, BVAE-TTS, in terms of speech quality. 
\end{abstract}

\section{Introduction}
\label{Introduction}
In recent years, text-to-speech (TTS) synthesis technologies have made rapid progress with deep learning techniques. Previous attention-based encoder-decoder TTS models have achieved state-of-the-art results in speech quality and intelligibility \cite{wang17, shen2018natural, gibiansky2017deep, ping2018deep, li2019neural}. They first generate acoustic feature (e.g., mel-spectrogram) autoregressively from text input using an attention mechanism, and then synthesize audio samples from the acoustic feature with a neural vocoder \cite{oord2016wavenet, kalchbrenner2018efficient}. However, the autoregressive (AR) structure has two major limitations: (1) it greatly limits the inference speed, since the inference time of AR models grows linearly with the output length; (2) the AR models usually suffer from robustness issues, e.g., word skipping and repeating, due to accumulated prediction error.

To avoid the aforementioned limitations of the AR TTS models, researchers have proposed various non-autoregressive (non-AR) TTS models \cite{ren2019fastspeech, peng2020non, ren2020fastspeech, kim2020glow, miao2020flow, lee2021bidirectional}. These models can synthesize acoustic features with significantly faster speed than AR models and reduce robustness issues, while achieving comparable speech quality to their AR counterparts. However, these models usually contain a separate duration module that does not propagate information to the acoustic module, which may lead to the training-inference mismatch issue. Moreover, the duration module needs duration labels as supervision, which may come from pre-trained AR TTS models \cite{ren2019fastspeech, peng2020non}, forced aligner \cite{ren2020fastspeech}, dynamic programming backtrack path \cite{kim2020glow} or jointly-trained attention modules \cite{miao2020flow, lee2021bidirectional}.


In attention-based encoder-decoder TTS models, the keys and values are calculated from text, while the queries are differently constructed in various models. 
For AR TTS models like Tacotron \cite{wang17}, the queries are constructed from the AR hidden states. It is a natural choice since the AR hidden states contain the acoustic information up to the current frame. For non-AR TTS models like FLOW-TTS \cite{miao2020flow}, only position information is used as query. 
For BVAE-TTS, the query is constructed form VAE latent variables.
If the correlation between the queries and keys are not strong enough, it is difficult for attention module to learn the alignment between queries and keys. Therefore, the key to build a non-AR TTS model is constructing the queries that are highly correlated to keys and also enable parallel attention calculation.


The corresponding text transcription can be considered as a lossy compression of the acoustic features. Meanwhile, the latent variables from hierarchical variational autoencoder (VAE) are also lossy compressions of acoustic features when trained on acoustic data. These latent variables may be correlated with the text and suitable to construct the queries. Moreover, hierarchical VAE provides latent variables of different resolutions, which can be used as the queries to build a coarse-to-fine layer-wise refined residual attention module. Specifically, the attention layer can use the alignment produced by the previous layer as an extra input and produce a refined version of the alignment. In this way, the burden of predicting the textual-to-acoustic alignment could be reduced and amortized by the coarse-to-fine attention layers. For the fist attention layer, we can use a ``nearly diagonal" initial alignment as a good instructive bias, considering the monotonic properties of text and acoustic features. 

In this work, we present a novel non-AR TTS model based on a specific hierarchical VAE model called Very Deep VAE (VDVAE) and a coarse-to-fine layer-wise refined residual attention module. 
Our main contributions are as follows:
\begin{itemize}
\item We adopt VDVAE to model mel-spectrograms with bottom-up and top-down paths. The hierarchical latent variables serve as queries to the attention modules.
\item We propose a novel residual attention mechanism that learns layer-wise alignments from coarse granularity to fine granularity in the top-down paths. 
\item We propose detailed Kullback-Leibler ($\operatorname{KL}$) gain for VDVAE to avoid posterior collapse in some hierarchical layers.
\item The proposed model is fully parallel and trained in an end-to-end manner, while obtaining high speech quality. Inference speed of the proposed model is 16 times faster than the AR model, Tacotron2, on a single NVIDIA GeForce RTX 2080 Ti GPU. Also, our proposed model outperforms an analogous non-AR model, BVAE-TTS, in terms of speech quality. 
\end{itemize}
The rest of the paper is organized as follows. Section \ref{sec:related_work} discusses the related work. The VAEs are introduced in Section \ref{sec:vae}. We present the model architecture in Section \ref{sec:model}. Experimental results and analyses are reported in Section \ref{sec:experiments} and Section \ref{sec:results}. The conclusion is drawn in Section \ref{sec:conclusion}.

\section{Related Work} \label{sec:related_work}
\subsection{Text-to-Speech Models}
The goal of text-to-speech (TTS) synthesis is to convert an input text sequence into an intelligible and natural-sounding speech utterance. 
Most previous work divides the task into two steps. The first step is text-to-acoustic (e.g. mel-spectrograms) modeling. 
Tacotron 1 \& 2 \cite{wang17, shen2018natural}, Deep Voice 2 \& 3 \cite{gibiansky2017deep, ping2018deep}, TransformerTTS \cite{li2019neural} and Flowtron \cite{valle2020flowtron} are AR models among the best performing TTS models. These models employ an encoder-decoder framework with attention mechanism, where the encoder convert the input text sequence to hidden representations and the decoder takes a weighted sum of the hidden representations to generate the output acoustic features frame by frame. 
It is challenging to learn the alignment between text sequence and acoustic features (e.g. mel-spectrograms) for TTS models. Various attention mechanisms are proposed to improve the stability and monotonicity of the alignment in AR models, such as location-sensitive attention \cite{chorowski2015attention}, forward attention \cite{zhang2018forward}, multi-head attention \cite{li2019neural}, stepwise monotonic attention \cite{he2019robust} location-relative attention \cite{battenberg2020location}. However, the low inference efficiency of AR models hinders their application in real-time services.
Recently, non-AR models are proposed to synthesize the output in parallel.
The key to design a non-AR acoustic model is the parallel alignment prediction. 
ParaNet \cite{peng2020non} also adopts a layer-wise refined attention mechanism, where the queries are only positional encoding in the first attention layer and previous attention layer output processed by a convolution block in the following attention layers. However, attention distillation from pre-trained AR TTS model are still needed to guide the training of alignments.
Fastspeech \cite{ren2019fastspeech} also requires knowledge distillation from pre-trained AR TTS model to learn alignments, while Fastspeech 2 bypasses the requirement of teacher model through an external force aligner for duration labels \cite{ren2020fastspeech}. Glow-TTS \cite{kim2020glow} and Flow-TTS \cite{miao2020flow} are both flow-based non-AR TTS models. Glow-TTS enforces hard monotonic alignments through the properties of flows and dynamic programming. Flow-TTS adopts positional attention to learn the alignment during training and uses length predictor to predict spectrogram length during inference. 

The second step is acoustic features to time-domain waveform samples modeling.  WaveNet \cite{oord2016wavenet} is the first of these AR neural vocoders, which produced high quality audio. Since WaveNet inference is computationally challenging. Several AR models are proposed to improve the inference speed while retaining quality \cite{arik2017deep, jin2018fftnet, kalchbrenner2018efficient}. Non-AR vocoders have also attracted increasing research interest \cite{oord2018parallel, ping2018clarinet, prenger2019waveglow, yamamoto2020parallel}, which generate high fidelity speech much faster than real-time.

Recently, end-to-end generation of audio samples from text sequence has been proposed in \cite{donahue2020end, ren2020fastspeech, weiss2020wave}. Wave-Tacotron extends Tacotron by incorporating a normalizing flow into the AR deocoder loop. 
Both EATS \cite{weiss2020wave} and Fastspeech 2s \cite{ren2020fastspeech} are non-AR models, where various adversarial feedbacks and auxiliary prediction losses are used respectively.


\subsection{VAE-based Generative Models}
VAE \cite{kingma2013auto, rezende2014stochastic} is a widely used generative model. Both variational RNN (VRNN) \cite{chung2015recurrent} and vector quantised-VAE (VQ-VAE) \cite{van2017neural} adopt AR structure to model the generative process of audio samples. 
\cite{child2020very} verifies that VDVAE outperforms the AR model PixelCNN \cite{van2016conditional} in log-likelihood on all natural image benchmarks, while using fewer parameters and generating samples thousands of times faster. Inspired by the success of the VDVAE architecture, we employ it for parallel speech synthesis task.

In parallel to our work, BVAE-TTS \cite{lee2021bidirectional} has been proposed, where bidirectional hierarchical VAE architecture is also adopted for non-AR TTS. However, only latent variables from the toppest layer are used as queries and there is a gap between the attention-based mel-spectrogram generation during training and the duration-based mel-spectrogram generation during inference in BVAE-TTS. Therefore, various empirical and carefully-designed techniques are needed to bridge this gap. By employing a residual attention mechanism, our proposed model can eliminate this training and inference mismatch. 

\section{Variational autoencoders} \label{sec:vae}
VAEs consist of the following parts: a generator $p(\mathbf{x}|\mathbf{z})$, a prior $p(\mathbf{z})$ and a posterior $q(\mathbf{z}|\mathbf{x})$ approximator.
Typically, the posteriors and priors in VAEs are assumed normally distributed with diagonal covariance, which allows for the Gaussian reparameterization trick to be used \cite{kingma2013auto}. The generator $p(\mathbf{x}|\mathbf{z})$ and approximator $q(\mathbf{z}|\mathbf{x})$ are jointly trained by maximizing the evidence
lower bound (ELBO):  
\begin{equation} \label{eq1}
\log p(\mathbf{x}) \geq \mathbb{E}_{\mathbf{z} \sim q(\mathbf{z}|\mathbf{x})} \log p(\mathbf{x}|\mathbf{z})-\operatorname{KL}\left[q(\mathbf{z}|\mathbf{x}) \| p(\mathbf{z})\right]
\end{equation}
where the first term of the right hand side of this inequality  can be seen as the expectation of negative reconstruction error and the second $\operatorname{KL}$ divergence term can be seen as a regularizer.

Hierarchical VAEs can gain better generative performance than previous VAEs where fully-factorized posteriors and priors are incorporated. One typical hierarchical VAE is introduced in \cite{sonderby2016ladder}, where both the prior $p(\mathbf{z})$ and posterior approximator $q(\mathbf{z}|\mathbf{x})$ are conditionally dependent:
\begin{equation} \label{eq2}
p(\mathbf{z})=p\left(\mathbf{z}_{0}\right) p\left(\mathbf{z}_{1}|\mathbf{z}_{0}\right) \ldots p\left(\mathbf{z}_{N}|\mathbf{z}_{<N}\right)
\end{equation}
\begin{equation} \label{eq3}
q(\mathbf{z}|\mathbf{x})=q\left(\mathbf{z}_{0}|\mathbf{x}\right) q\left(\mathbf{z}_{1}|\mathbf{z}_{0}, \mathbf{x}\right) \ldots q\left(\mathbf{z}_{N}|\mathbf{z}_{<N}, \mathbf{x}\right)
\end{equation}

\begin{figure*}[t]
	\centering
	\includegraphics[width=16cm]{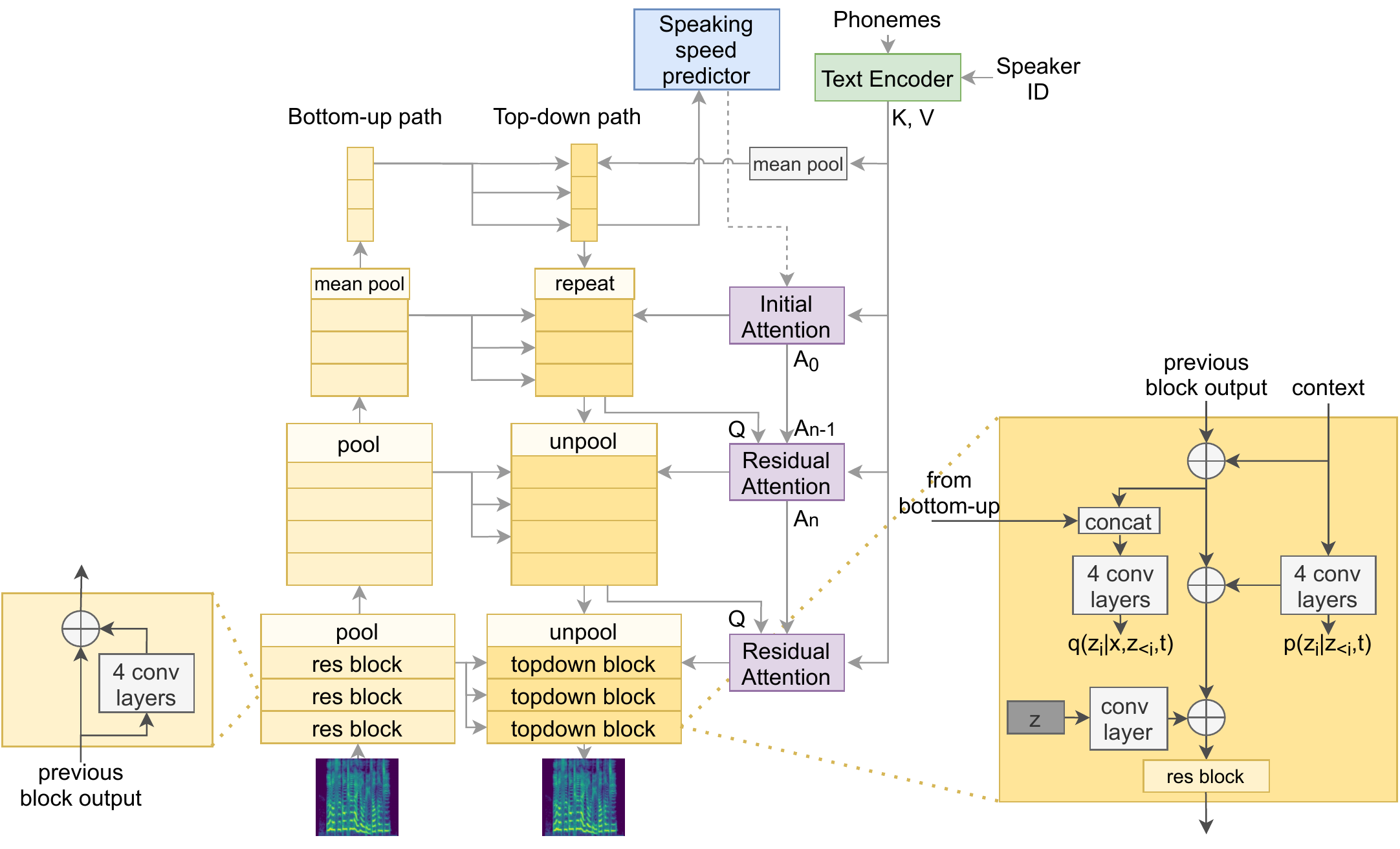}	
	\vspace{-0.25cm}
	\caption{Overall model architecture}
	\label{model}
\end{figure*}

where N is the number of hierarchical groups. This hierarchical VAE first extracts latent variables from input data $\mathbf{x}$ along the bottom-up path, then processes the latent variables along the top-down path to generate $\mathbf{x}$. The equation (\ref{eq1}) is changed as follows:

\begin{equation} \label{eq4}
\begin{split}
\log &p(\mathbf{x}) 
 \geq \mathbb{E}_{\mathbf{z} \sim q(\mathbf{z}|\mathbf{x})} \log p(\mathbf{x}|\mathbf{z})-\operatorname{KL}\left[q(\mathbf{z}_{0}|\mathbf{x}) \| p(\mathbf{z}_{0})\right] \\
& -\sum_{n=1}^{N} \mathbb{E}_{q\left(\mathbf{z}_{<n}|\mathbf{x}\right)} \left[
\operatorname{KL}\left[q\left(\mathbf{z}_{n}|\mathbf{z}_{<n},\mathbf{x} \right) \| p\left(\mathbf{z}_{n}|\mathbf{z}_{<n}\right)\right)]\right]
\end{split}
\end{equation}

where $q\left(\boldsymbol{z}_{<n}|\boldsymbol{x}\right):=\prod_{i=1}^{n-1} q\left(\boldsymbol{z}_{i}|\boldsymbol{z}_{<i}, \boldsymbol{x} \right)$ is the approximate posterior up to the $(n-1)^{t h}$ group.

\section{Model Architecture} \label{sec:model}
We adopt the VDVAE \cite{child2020very} with a novel residual attention mechanism for non-AR TTS. The overall architecture is shown in Figure \ref{model}. 
Hierarchical latent variables at decreasing time scale are extracted from the input mel-spectrograms along the bottom-up path. These hierarchical latent variables are processed from top to bottom and served as queries ($\mathbf{Q}$). A text encoder takes phoneme sequence and optional speaker ID as input and outputs text encoding as key ($\mathbf{K}$) and value ($\mathbf{V}$). $\mathbf{Q}$, $\mathbf{K}$, $\mathbf{V}$ and attention weight from previous attention block ($\mathbf{A}_\text{prev}$) are sent to the following residual attention module to produce a refined attention weight and context vector. The context vector is a weighted average of $\mathbf{V}$, and is passed to top-down block for variational inference and mel-spectrogram reconstruction along with stored hierarchical latent variables. A speaking speed predictor takes the mean-pooled latent variables of the coarsest layer as input and predicts the utterance-level average speaking speed factor to determine the number of acoustic frames at inference.

We use the $\beta$-VAE training objective and a mel-spectrogram loss inspired by \cite{rezende2018tamingvaes}. 
The hierarchical VAE, speaking speed predictor, text encoder and residual attention modules are jointly trained with the following objective:


\begin{equation} \label{eq5}
    \mathcal{L} = \alpha\mathcal{L}_{\text{speaking\_speed}} + \mathcal{L}_{\text{recon}} +\beta\mathcal{L}_{\operatorname{KL}} 
\end{equation}
where,
\begin{equation} \label{eq6}
    \mathcal{L}_{\text{speaking\_speed}}=\mathbb{E}\left(\mathrm{d}-\hat{\mathrm{d}}\right),
\end{equation}
\begin{equation} \label{eq7}
\mathcal{L}_{\text{recon}}=\frac{\left\|\mathbf{x}-\mathbf{\hat{x}}\right\|_{2}}{\left\|\mathbf{x}\right\|_{2}}+\left\|\log \mathbf{x}-\log \mathbf{\hat{x}}\right\|_{1},
\end{equation}
\begin{equation} \label{eq8}
\begin{split}
    &\mathcal{L_{\operatorname{KL}}}=\operatorname{KL}\left[q(\mathbf{z}_{0}|\mathbf{x},\mathbf{t}) \| p(\mathbf{z}_{0}, \mathbf{t})\right] +  \\
    & \sum_{n=1}^{N} \mathbb{E}_{q\left(\mathbf{z}_{<n}|\mathbf{x}, \mathbf{t}\right)}\left[\operatorname{KL}\left[q\left(\mathbf{z}_{n}|\mathbf{x}, \mathbf{z}_{<n}, \mathbf{t}\right) \| p\left(\mathbf{z}_{n}|\mathbf{z}_{<n}, \mathbf{t}\right)\right]\right],
\end{split}
\end{equation}
and $\alpha$, $\beta$ are weights of speaking rate loss, $\operatorname{KL}$ loss and respectively. $\mathrm{d}$, $\hat{\mathrm{d}}$, $\mathbf{x}$, $\mathbf{\hat{x}}$, $\mathbf{t}$ and $\mathbf{z}$ denote ground-truth speaking speed, predicted speaking speed, ground-truth mel-spectrogram magnitudes, predicted mel-spectrogram magnitudes, phoneme sequences, and latent variables respectively. 
We introduce detailed implementations of VDVAE, text encoder, speaking rate predictor and residual attention modules in the following subsections.

\subsection{Very deep VAE}
As show in the central part of Figure \ref{model}, the very deep VAE is stacked of residual block groups along the bottom-up path and topdown block groups along the top-down path. 
The time dimension of input mel-spectrogram is reduced at the pooling layer of each bottom-up group using average pooling. The residual block contains four convolution layers and a residual connection as shown in the left part of Figure \ref{model}. Each convolution output is preceded by the GELU nonlinearity \cite{hendrycks2016gaussian}. 
The topdown block takes bottom-up output, previous topdown block output and context vector from residual attention module as input. The previous topdown block output is unpooled at the bottom layer of each top-down group using nearest-neighbor upsampling along temporal axis, and then added to context vector as block bias for prior $p(\cdot)$ and the posterior approximator $q(\cdot)$ . 
The bottom-up group output is concatenated with this block bias and processed by four convolution layers to predict $q\left(\boldsymbol{z}_{i}|\boldsymbol{x}, \boldsymbol{z}_{<i}, \boldsymbol{t}\right)$. Another path of context vector is used to predict the prior $p\left(\boldsymbol{z}_{i}|\boldsymbol{z}_{<i}, \boldsymbol{t} \right)$ through another four convolution layers. Both $q(\cdot)$ and $p(\cdot)$ are isotropic Gaussian distributions. 
In training stage, $\boldsymbol{z}$ is sampled from $q(\cdot)$. In inference stage, the bottom-up path is discarded and $\boldsymbol{z}$ is sampled from $p(\cdot)$. 
$\boldsymbol{z}$ passes a convolution layer and is added to the summation of the block bias for posterior approximator and the output of the four convolution layers, before being sent to a residual block. The residual block contains three convolution layers, one dilated convolution layer and a residual connection. $\boldsymbol{z}$ is also used as the query in the succeeding residual attention module.

Although residual connections in VDVAE are able to alleviate posterior collapse to some degree, we observe some horizontal segments in the cumulative $\operatorname{KL}$ curve as shown in Figure \ref{fig:cum_kl}.
This indicates that posterior collapse occurs in these layers and latent variables from these layers encode small amount of information. Therefore, we propose a detailed $\operatorname{KL}$ gain mechanism, inspired by \cite{alemi2018elbo,burgess2018betavae}.
In hierarchical VAE, the $\operatorname{KL}$ term is the accumulation of the $\operatorname{KL}$ terms from different hierarchical layers. We can set a reference value for each layer like the $\operatorname{KL}$ reference term in \cite{alemi2018elbo}:
\begin{equation} \label{rate_gain}
    \operatorname{KL}_{\text{ref}} = \frac{c}{N} \sum_{i=1}^{N} \operatorname{KL}_i
\end{equation}
\begin{equation} \label{10}
    \mathcal{L}_{\text{detailed\_KL\_gain}}= \sum_{i=1}^{N} {\left| { \max { \left(  \operatorname{KL}_i, \operatorname{KL}_{\operatorname{ref}} \right) } - \operatorname{KL}_{\text{ref}} }\right| }
\end{equation}

where the reference $\operatorname{KL}_{ref}$ is the average value of $\operatorname{KL}$ terms from different hierarchical layers multiplied by a constant $\operatorname{KL}$ gain factor $c$ ($c=0.5$ is used in this paper). When $\operatorname{KL}_i$ is larger than the reference, the $\operatorname{KL}$ gain term is $0$. When $\operatorname{KL_i}$ is smaller than the reference, the gain term is non-zero and the gradient of this term would enlarge the $\operatorname{KL_i}$, then the ineffective latent variables would be eliminated. Therefore, the training objective of the whole model is modified as follows:
\begin{equation} \label{eq11}
    \mathcal{L} = \alpha\mathcal{L}_{\text{speaking\_speed}} + \mathcal{L}_{\text{recon}} +\beta\mathcal{L}_{\operatorname{KL}} + \lambda\mathcal{L}_{\text{detailed\_KL\_gain}}
\end{equation}
where $\lambda$ is the weight of the detailed $\operatorname{KL}$ gain.

\subsection{Speaking speed predictor}
Since number of mel-spectrogam frames to generate during inference is unknown a priori, we train an auxiliary speaking speed predictor.
A two-dimensional mel-spectrogram is reduced in time-scale hierarchically along the bottom-up path and eventually to a vector $\mathbf{x}_0$ by a temporal global mean pooling operation; another vector $\mathbf{t}_0$ is obtained from the text encoder output by temporal global mean pooling.
$\mathbf{x}_0$ and $\mathbf{t}_0$ are used to calculate the prior and posterior distributions for the top-most latent variables $\mathbf{z}_0$.
We sample a latent variable vector $\mathbf{z}_0$ from the latent space, which is fed into the speaking speed predictor as input. During training, $\mathbf{z}_0$ is sampled from the posterior $q(\mathbf{z}_0|\mathbf{x}_0, \mathbf{t}_0)$, while at inference, $\mathbf{z}_0$ is sampled from the prior $p(\mathbf{z}_0|\mathbf{t}_0)$. 

According to the InfoGAN derivation \cite{chen2016infogan}, minimizing the prediction error for speaking speed from the top-most latent variables paves a way to maximizing a lower bound of their mutual information (proof in Appendix A). If the speaking speed predictor is well-trained, latent variables from the first top-down group (i.e., the top-most latent variable $\mathbf{z}_0$) would encode rich speaking speed information.
Since latent variables (the $\mathbf{z}$'s) in VARA-TTS are conditionally dependent in a chain from top-to-bottom, they also contain speaking speed information, which facilitates the alignment learning at their corresponding attention modules.  
This alleviates the problem of inconsistency of acoustic model and a separately trained duration model.  
For example, the duration module of BVAE-TTS does not propagate information to the acoustic module and only uses the alignment obtained from the acoustic module and text encoding module as supervision. This may lead to the issue that the acoustic module is not adapted well to the duration module.

In VARA-TTS, the speaking speed predictor adopts two fully-connected (FC) layers, between which we add a ReLU activation function, a layer-normalization layer and a dropout layer in a sequence (more details are presented in Appendix D). Unlike most existing non-AR TTS models, which require token-level duration information as supervision for duration model training, we use a readily-computed speaking rate $\mathrm{d}$ for each training utterance as the target of the speaking speed predictor:
\begin{equation}
    \mathrm{d} = Normalize_\text{min-max}(\frac{T_{\text{mel}}}{L_{\text {text}}}) \in [0, 1],
\end{equation}
where $Normalize_\text{min-max}(\cdot)$ denotes min-max normalization across the training set, $T_{mel}$ represents the number of frames in a mel spectrogram and $L_{\text{text}}$ is the number of tokens in a phoneme sequence. We add a sigmoid function on the speaking speed predictor output to obtain the predicted speaking rate $\hat{\mathrm{d}}$. An MSE loss applied on $\mathrm{d}$ and $\hat{\mathrm{d}}$ is used as training signal for the speaking speed predictor.


\begin{figure}[t]
    \centering
	\includegraphics[width=5cm]{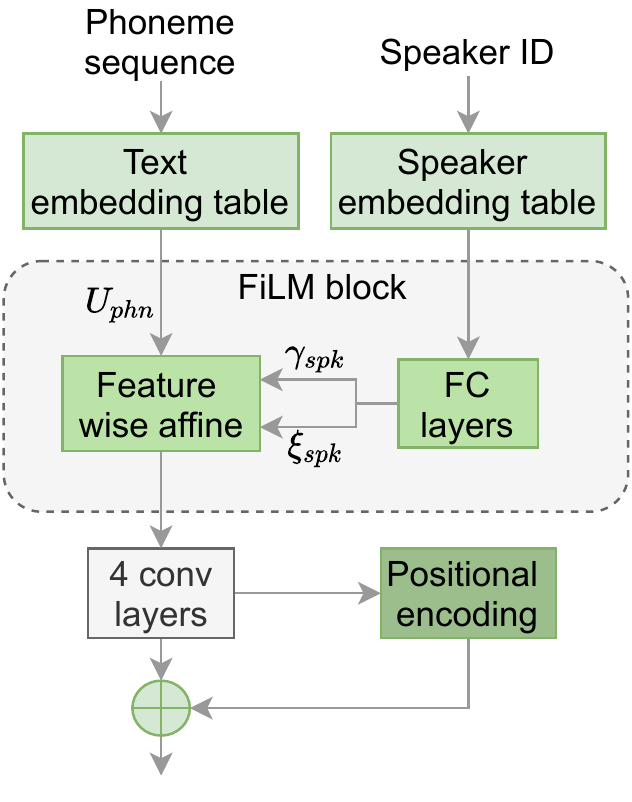}		
	\vspace{-0.45cm}
	\caption{Network detail of text encoder}
	\label{text_enc}
\end{figure}

\subsection{Text encoder}
As shown in Fig \ref{text_enc}. phoneme sequence and speaker id are transformed to phoneme embedding and speaker embedding through two lookup tables. The feature-wise linear modulation (FiLM) \cite{perez2018film} is used to fuse speaker embedding with phoneme embedding. Two FC layers are used to compute the scale and shift vectors from the speaker embedding vector, respectively. The feature-wise affine operation is conducted as:
\begin{equation} \label{12}
\gamma_{\text {spk}} \times U_{\text {phn }} + \xi_{\text {spk }}
\end{equation}
where $\gamma_{\text {spk}}$ and $\xi_{\text {spk }}$ represent the scale and shift vectors and $U_{\text {phn }}$ represents phoneme embedding.
The fused output passes four convolution layers. The convolution output and sinuous positional encoding \cite{vaswani2017attention} are added as text encoder output.

\subsection{Residual attention}
For the initial attention module in Figure \ref{model}, we simply set the attention weight matrix $\mathbf{A} \in \mathbb{R}^{T_{\text{max\_red}} \times L_{\text{text}}}$ to be ``nearly diagonal". 
$T_{\text{max\_red}}$ is the number of feature frames before the mean pooling layer at bottom-up path. 
This ``nearly diagonal" attention matrix is a good instructive bias, since the alignment between text and acoustic features are monotonic and almost diagonal. 
Inspired by \cite{tachibana2018guidedattention}, we set the attention weight matrix as follows:
\begin{equation} \label{eq13}
S_{l} =\exp \left[-(t/T_{\text{max\_red}} - l/L_{\text{text}}) / 2 g^{2}\right]
\end{equation}
\begin{equation} \label{eq14}
A_{tl}= S_{tl} / \sum_{{l}^{\prime}} S_{tl^{\prime}}
\end{equation}
where, $l$ is text position index and $t$ is mel-spectrogram position index. In this paper, we generate four attention matrices using $g \in [0.01, 0.05, 0.1, 0.2]$ and compute four context vectors by multiplying the four generated attention matrices with text encoder output respectively. Then the four context vectors are concatenated and projected to one context vector along temporal axis. 
During training stage, $T_{\text{max\_red}}$ is obtained from the number of ground-truth mel-spectrogram frames divided by the maximum reduction factor. Maximum reduction factor is the product of the reduction factors between neighbouring residual block groups.   
During inference stage, it is obtained by multiplying predicted speaking speed $\hat{\mathrm{d}}$ to $L_{\text{text}}$ and then being divided by the maximum reduction factor. 

For the remaining residual attention modules, they take text encoder output ($\mathbf{K}$, $\mathbf{V}$) , attention weight from previous attention module ($\mathbf{A}_\text{prev}$) and latent variables from previous topdowm group ($\mathbf{Q}$) as input. Multi-head attention mechanism \cite{vaswani2017attention} is used, where $\mathbf{A}_\text{prev}$ is added as an additional input:
\begin{equation} \label{eq15}
\begin{array}{c}
\text {ResidualMultiHead} \left (\mathbf{Q}, \mathbf{K}, \mathbf{V},  \mathbf{A}_\text{prev} \right)= \\
\qquad \text {Concat}\left(\mathbf{head}_{1}, \ldots, \mathbf{head}_{h}\right) \mathbf{W}^{O}
\end{array}
\end{equation}
where $\mathbf{Q}$, $\mathbf{K}$, $\mathbf{V}$ are matrices with dimension $d_{k}$, $d_{k}$ and $d_{v}$ respectively. 
$\mathbf{head}_{i}$ = $\text{ResidualAttention}$ $(\mathbf{QW}_{i}^{Q}, \mathbf{KW}_{i}^{K},$ $\mathbf{VW}_{i}^{V}$
, $\mathbf{A}_\text{prev})$. 
$\mathbf{W}^{O}$ is the transformation matrix that linearly projects the concatenation of all heads output.
$\mathbf{W}_{i}^{Q}$, $\mathbf{W}_{i}^{K}$ and $\mathbf{W}_{i}^{V}$ are projection matrices of the $i$-th head. $\mathbf{A}_\text{prev}$ is the averaged attention weights from previous attention heads. We use the scaled dot-product mechanism based residual attention:
\begin{equation} \label{eq16}
\begin{array}{l}
\text {ResidualAttention}\left(\mathbf{Q}, \mathbf{K}, \mathbf{V},  \mathbf{A}_\text{prev} \right)= \\
\qquad \text{Softmax}\left(\frac{\mathbf{Q} \mathbf{K}^{T}}{\sqrt{d_{k}}}+\mathbf{A}_\text{prev}\right) \mathbf{V}
\end{array}
\end{equation}
We find that using attention weights $\mathbf{A}_\text{prev}$ makes training process more stable than using before-softmax attention scores, which is used in RealFormer \cite{he2020realformer}. The previous attention weight $\mathbf{A}_\text{prev}$ is first upsampled to fit the time dimension and then processed by a convolution layer before being sent to the next layer. Different from the location sensitive attention \cite{chorowski2015attention} that only takes into account attention weight at previous decoder time step, our residual attention mechanism incorporates attention weights of all time steps from last residual attention module, which provides a global perspective of attention history.    

\section{Experiments} \label{sec:experiments}

\subsection{Data}

We conduct both single-speaker and multi-speaker TTS experiments to evaluate the proposed VARA-TTS model. For single-speaker TTS, the LJSpeech corpus~\cite{ljspeech17} is used, which contains 13100 speech samples with total duration of about 24 hours. We up-sample the speech signals from 22.05 kHz to 24 kHz in sampling rate. The dataset is randomly partitioned into training, validation and testing sets according to a 12900/100/100 scheme. For multi-speaker TTS, we use an internal Mandarin Chinese multi-speaker corpus, which contains 55 hours of speech data from 7 female speakers.

In all experiments, mel-spectrograms are computed with 1024 window length and 256 frame shift. We convert text sentences into phoneme sequences
with Festival for English and with an internal grapheme-to-phoneme toolkit for Chinese.

\subsection{Model configurations}

We compare VARA-TTS with a well-known AR TTS model, Tacotron 2~\cite{shen2018natural}, and an analogous non-AR TTS model, BVAE-TTS~\cite{lee2021bidirectional}, under single-speaker TTS setting.
We use an open-source Tacotron 2 implementation\footnote{\url{https://github.com/NVIDIA/tacotron2}} and the official BVAE-TTS implementation\footnote{\url{https://github.com/LEEYOONHYUNG/BVAE-TTS}}. Some key hyperparameters in VARA-TTS are presented in Appendix C. To make the comparison fair, we use the same neural vocoder, HiFi-GAN~\cite{kong2020hifi}
with HiFi-GAN-V1 configuration, to convert mel spectrograms into waveform for all three compared models. 

We train the VARA-TTS model with a batch size of 32 with two NVIDIA V100 GPUs. 
And the model for evaluation are trained for 90k iterations. The Adam optimizer with $\beta_1=0.9$, $\beta_2=0.999$ is adopted for parameter updating, where the maximum of learning rate is 1.5e-4 and scheduled in the same manner as in \cite{vaswani2017attention} with 10k steps to warm-up.

\section{Results and Analysis} \label{sec:results}

\subsection{Speech quality}
Mean opinion score (MOS) test is conducted to evaluate the speech naturalness. We invite 10 raters to take the MOS test, where participants are asked to give a score from 1 to 5 (least natural to most natural) with 0.5 point increments to each stimuli presented. The sentences for evaluation are sampled from LibriTTS~\cite{zen2019libritts} test set. The MOS test results are presented in Table~\ref{table:mos_infspeed}. We can see that our proposed VARA-TTS obtains higher MOS score than BVAE-TTS, while is inferior to Tacotron 2 in naturalness. We also conduct a MOS test for a multi-speaker VARA-TTS trained with the multi-speaker Mandarin corpus. The MOS result is 4.49±0.11. We can see that the MOS score on multi-speaker Mandarin corpus is much higher than the result on LJSpeech. One possible reason may be that the multi-speaker Mandarin corpus is of higher quality and contains larger amount of data. VARA-TTS is data-hungry and tends to over-fit on LJSpeech.
Audio samples can be found online\footnote{\url{https://vara-tts.github.io/VARA-TTS/}}.

\subsection{Inference speed}
Benefiting from its non-AR structure, VARA-TTS enjoys fast inference speed.
We use the test set to compare the inference speed of the three compared models. 
Average inference speed of the three compared models obtained from 10 different runs on one NVIDIA GeForce RTX 2080 Ti GPU is presented in Table~\ref{table:mos_infspeed}. We can see that inference speed of our VARA-TTS is 16x faster than that of Tacotron 2, and at the same scale as that of BVAE-TTS.

\subsection{Alignment refinement process}
Figure B.1 in Appendix B shows an example of the alignment refinement process for an utterance by the well-trained multi-speaker VARA-TTS model.
In the figure, the bottom row shows the initial alignments, which show diagonal patterns generated by rule. In the second row, a blur alignment is generated. The blurriness indicates that the alignment is not reliable. However, VARA-TTS learns to refine the coarse alignment and obtain clearer alignments during the succeeding hierarchies, as can be seen from the upper plots in Figure B.1.
Moreover, we observe that the alignment refinement process varies with the $\beta$ value used.
For $\beta=1.0$, the alignments disperse in upper layers, and remain clear for $\beta=1.8$.
This indicates that hierarchical latent variables from the same layers encode different information when trained with different values of $\beta$.

\subsection{Ablation studies}
To show the effectiveness of model design in VARA-TTS, we conduct the following ablation studies: (i) Train without detailed $\operatorname{KL}$ gain; (ii) Use different values of $\beta$ in Equation \ref{eq11} during training; (iii) Separately train the speaking speed predictor.

\begin{figure}[t]
    \centering
	\includegraphics[width=5cm]{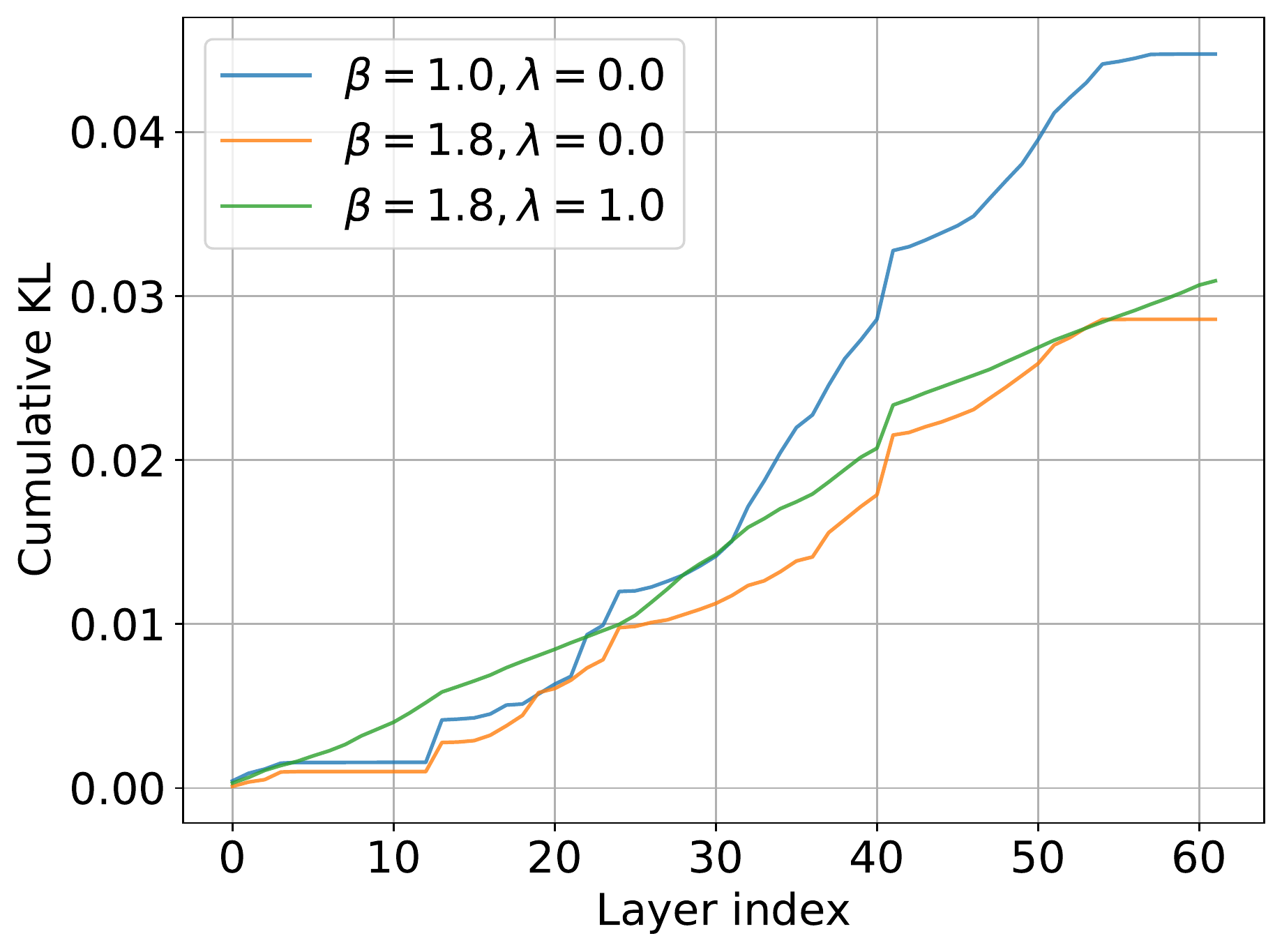}	
	\vspace{-0.35cm}
	\caption{Cumulative $\operatorname{KL}$ for different $\beta$ and $\lambda$. The y-axis is the cumulative $\operatorname{KL}$ value per data point by layers and the y-axis is the layer index. Flat horizontal line segment indicates that posterior collapse occurs in the corresponding layers.}
	\label{fig:cum_kl}
\end{figure}

\textbf{Detailed $\mathrm{KL}$ gain}. Figure~\ref{fig:cum_kl} shows the cumulative $\operatorname{KL}$ divergence curves with different values of $\beta$ and $\lambda$. We can see that there are some horizontal segments in the cumulative $\operatorname{KL}$ divergence curve when $\lambda=0$. This indicates $\operatorname{KL}$ values are $0$ and posterior collapse occurs in the corresponding layers.
When detailed $\operatorname{KL}$ gain is applied ($\lambda=1.0$), the cumulative $\operatorname{KL}$ increases smoothly with the layer index and contains no horizontal segments. The $\operatorname{KL}$ value of $\lambda=1.0$ is larger than that of $\lambda=0$ in Figure~\ref{fig:cum_kl}. This is because detailed $\operatorname{KL}$ gain mechanism enlarges the hierarchical $\operatorname{KL}$ values smaller than $\operatorname{KL}_\text{ref}$, which can be compensated by a larger $\beta$.
We also observe that when trained with detailed $\operatorname{KL}$ gain, VARA-TTS can learn clearer alignments in coarse hierarchical layer as shown in Appendix B. This indicates that posterior collapse does not happen in these layers and the latent variables encode meaningful information.
\begin{table}[]
\centering
\caption{MOS with 95\% confidence and inference speed results.}
\begin{tabular}{l|c|c}
\hline
Model & MOS & Inference speed (ms) \\ \hline
Tacotron 2 & 4.11±0.22 & 526.52 \\ \hline
BVAE-TTS & 3.33±0.18 & 18.06 \\ \hline
VARA-TTS & 3.88±0.20 & 32.01 \\ \hline
\end{tabular}
\label{table:mos_infspeed}
\end{table}

\begin{figure*}[h]
        \centering
          \subfloat[\centering feature reconstruction error]{%
              \includegraphics[height=3.8cm]{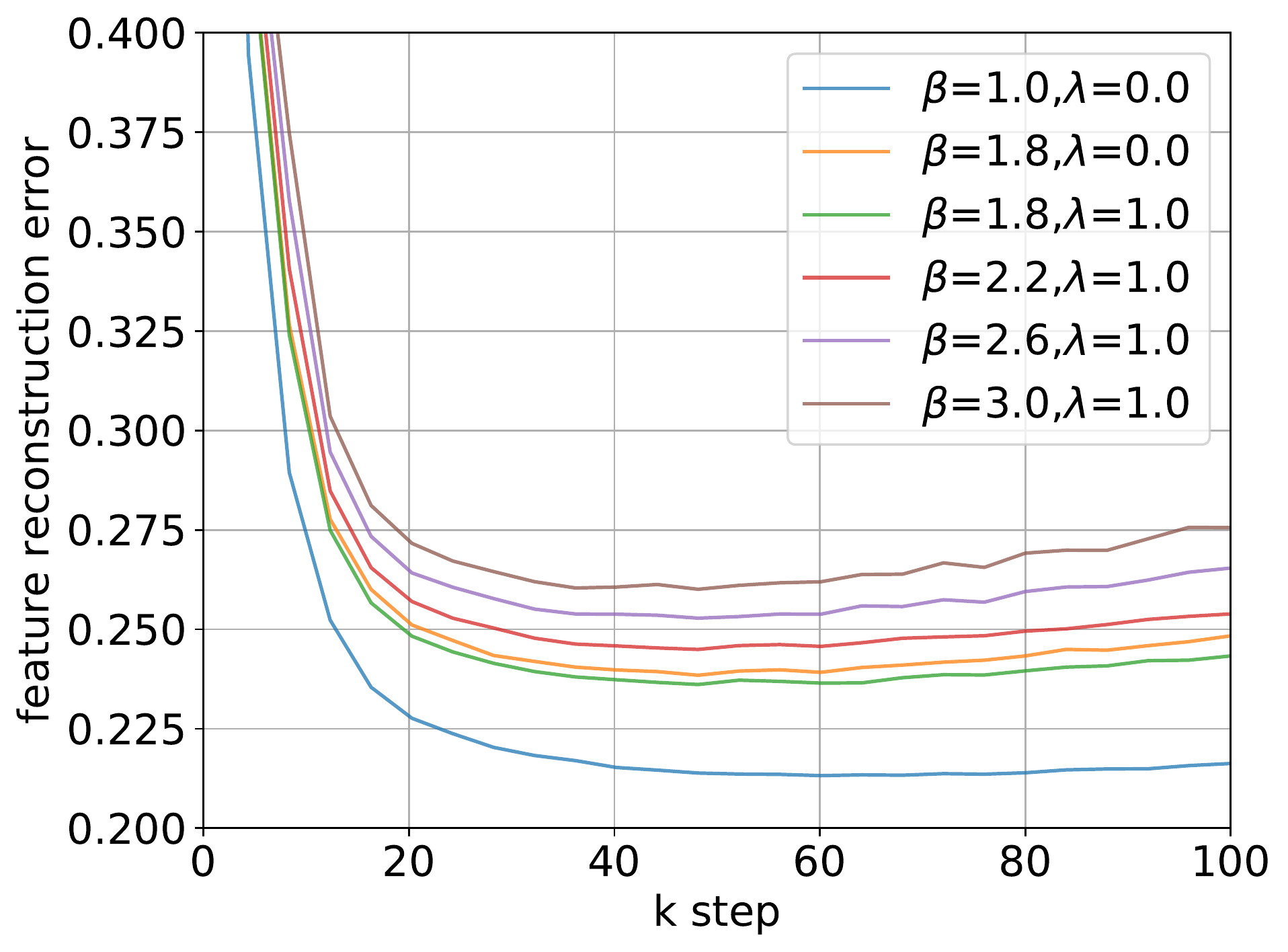}%
              \label{fig:distortion_1}%
          } 
          \qquad
          \subfloat[\centering $\operatorname{KL}$ ]{%
              \includegraphics[height=3.8cm]{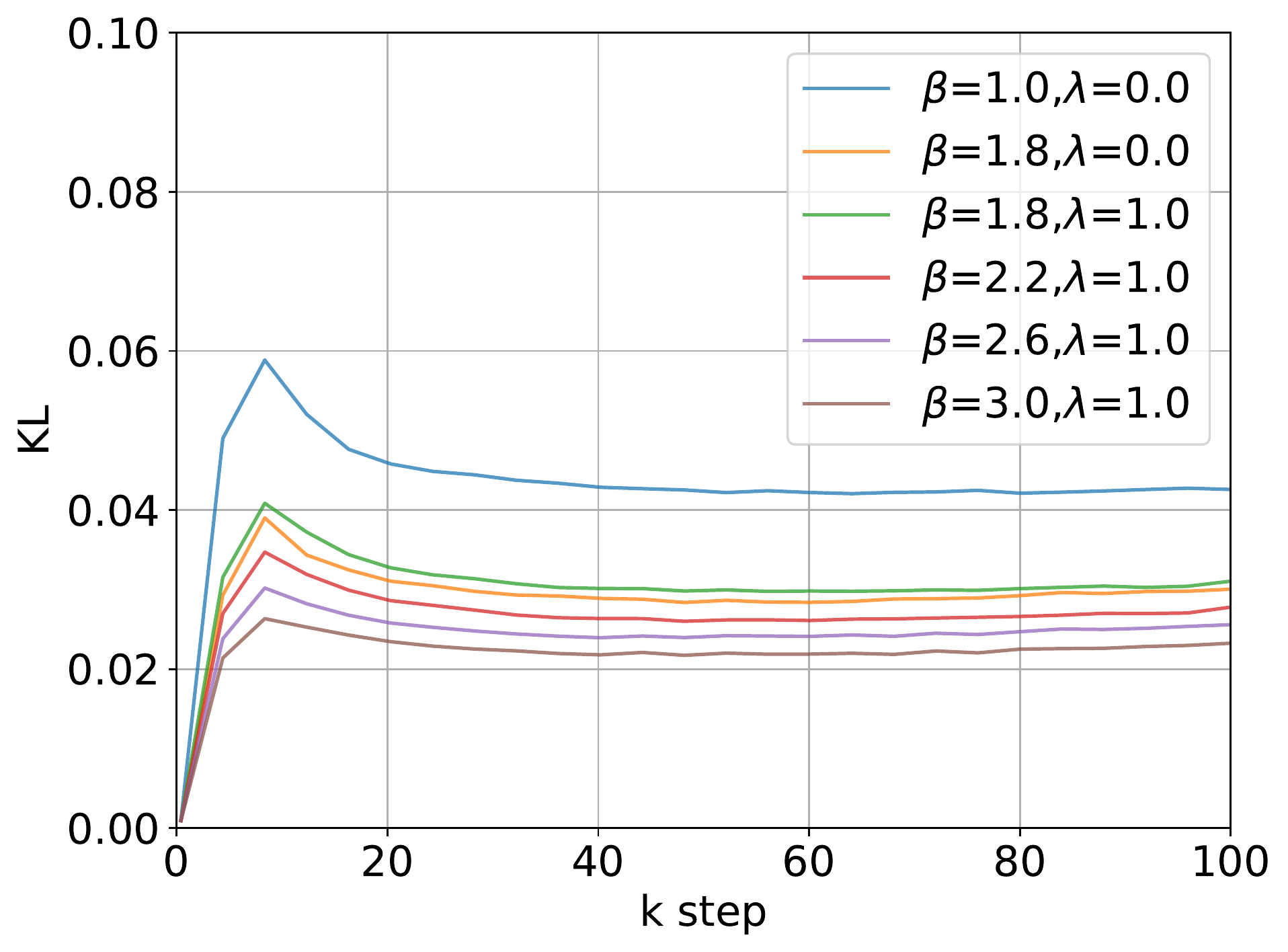}%
              \label{fig:rate_2}%
          }
          \qquad
          \subfloat[\centering speaking speed error]{%
              \includegraphics[height=3.8cm]{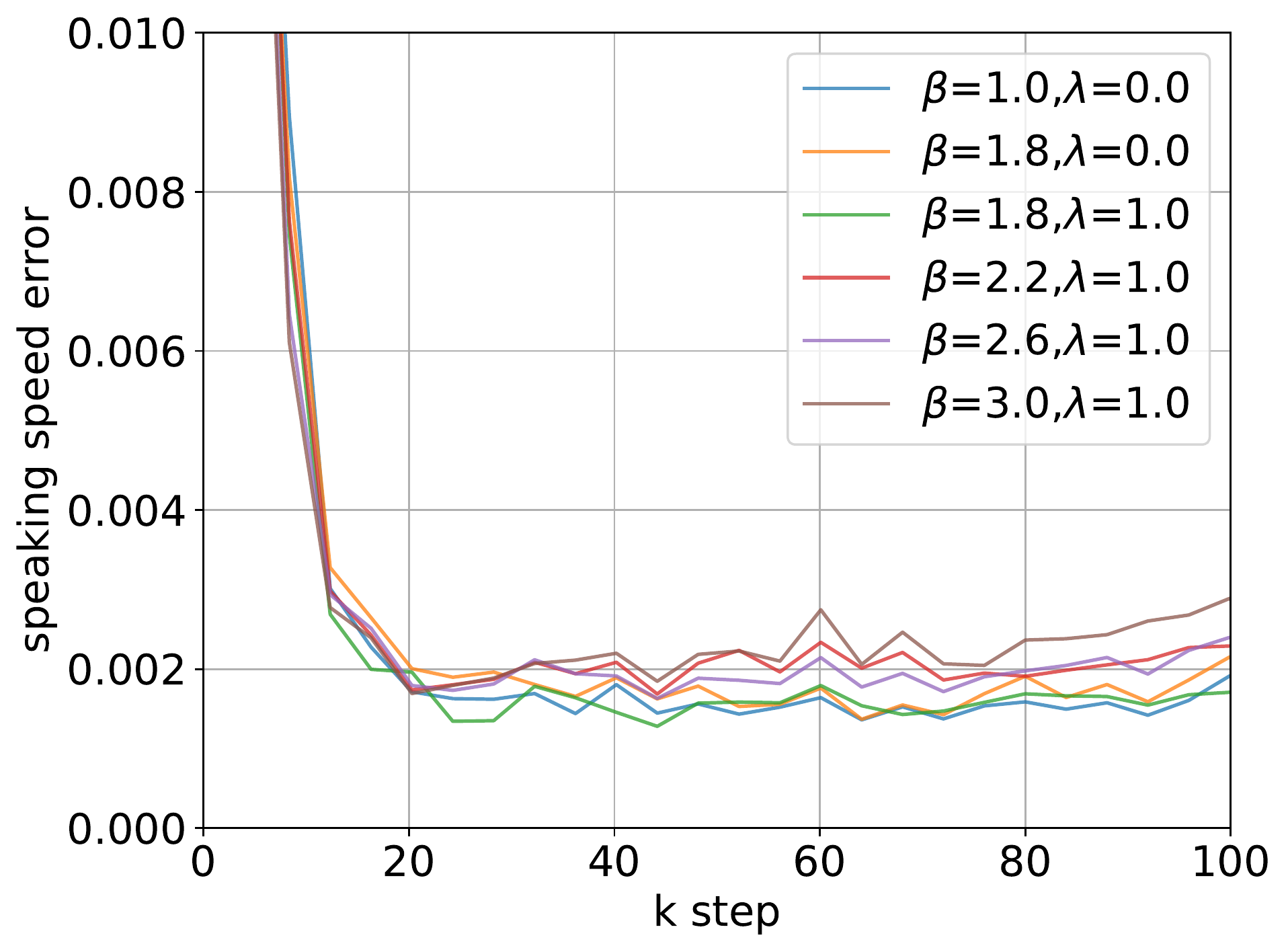}%
              \label{fig:dur_loss_4}%
          }
          \vspace{-0.2cm}
          \caption{The feature reconstruction error, $\operatorname{KL}$, and speaking speed error training curves on validation set with
          different values of $\beta$ and $\lambda$. The x-axis is training step in the unit of thousand and the y-axis is the corresponding value of feature reconstruction error, $\operatorname{KL}$, and speaking speed error.}
          \label{fig:dist_rate}
\end{figure*} 

\begin{figure*}[t] 
\centering
\subfloat[\centering speaking speed error]{%
	\includegraphics[width=5cm]{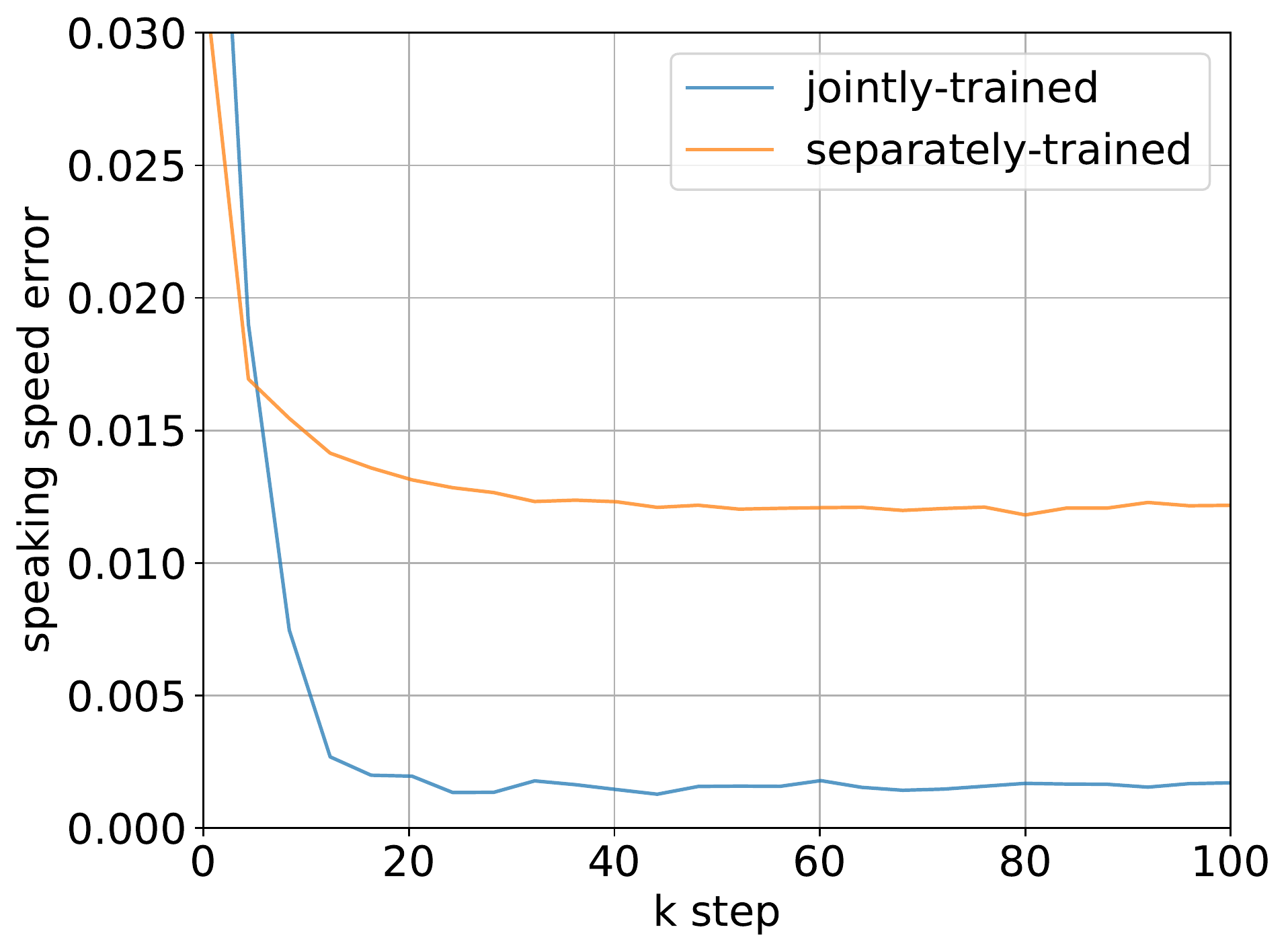}		
	\label{fig:dur_dur_loss}
} 
\qquad
\subfloat[\centering -ELBO]{%
	\includegraphics[width=5cm]{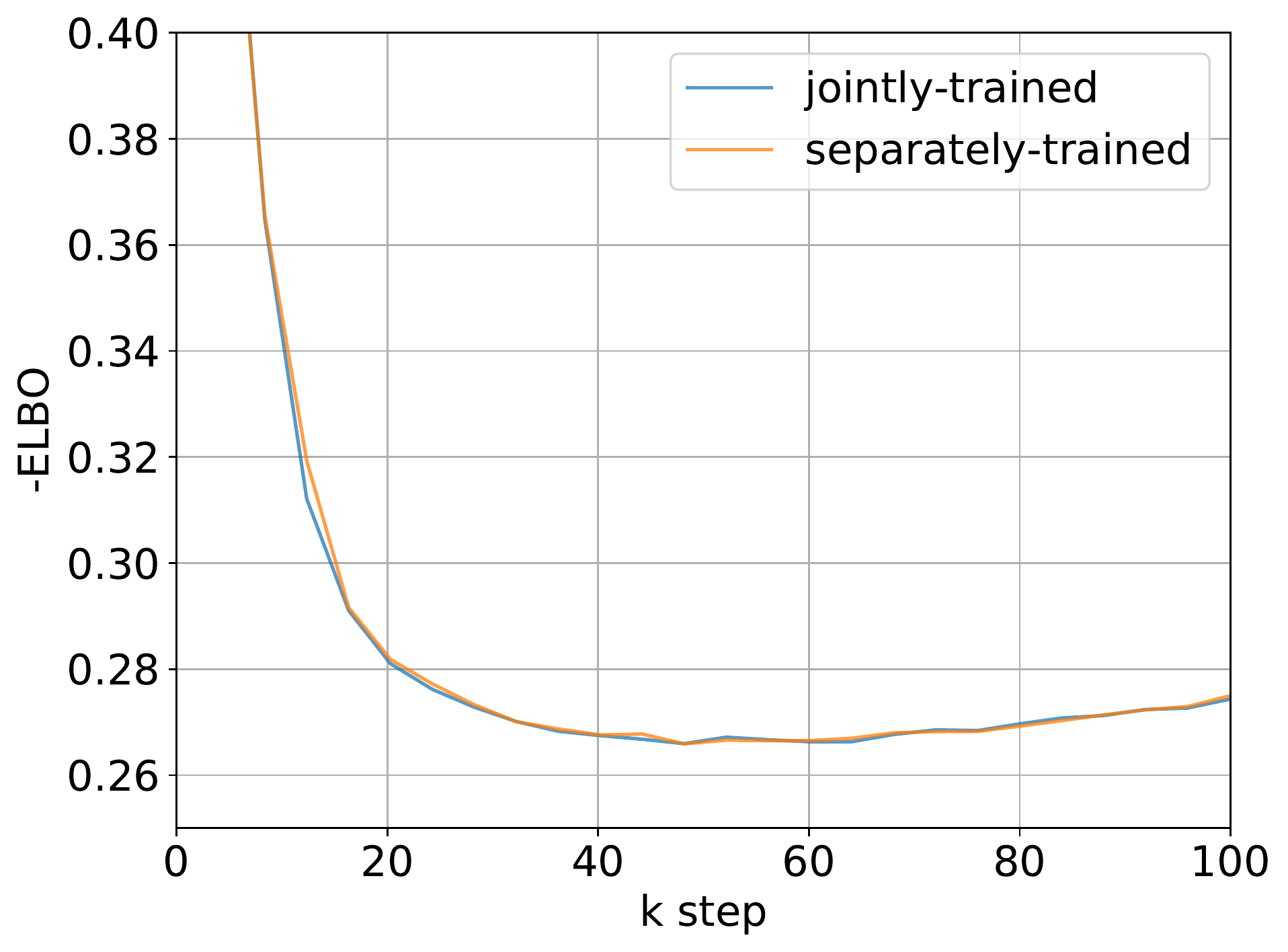}		
	\label{fig:dur_elbo_loss}
}
\vspace{-0.2cm}
\caption{Speaking speed error and -ELBO for models trained with a joint speaking speed predictor and a separate one}
\label{fig:dur_loss}
\end{figure*}

\textbf{Different values of $\mathbf{\beta}$}. At training, the queries are sampled from posterior, but sampled from prior at inference.
Smaller $\operatorname{KL}$ value indicates smaller gap between prior and posterior for VDVAE.
The smaller the gap between prior and posterior, the more reliable the queries can be at inference.
The hpyerparameter $\beta$ in Equation \ref{eq11} adjusts the relative weight between the reconstruction error and $\operatorname{KL}$ term. The curves of feature reconstruction error and $\operatorname{KL}$ value with different values of $\beta$ and $\lambda$ on validation set are shown in Figure \ref{fig:dist_rate} (a) and (b) respectively.
As can be seen, enlarging $\beta$ results in larger feature reconstruction error and smaller $\operatorname{KL}$ value.
However, we find that small changes in feature reconstruction error do not influence perpetual result a lot.
Figure \ref{fig:dist_rate} (c) shows the speaking speed error on validation set. 
We can see that different values of $\beta$ have little effect on speaking speed error.
We use $\beta=1.8$ and $\lambda=1.0$ for model evaluation.

\textbf{Joint speaking speed modeling}. The separate speaking speed predictor has the same structure as the joint one, 
except that its input is the mean-pooled text embedding and detached from the computational graph by a stop gradient operation. 
The stop gradient operation is also applied in BVAE-TTS and GLOW-TTS to avoid affecting the training objective.
As show in Figure \ref{fig:dur_loss}, the joint training strategy obtains similar ELBO as the separate training counterpart, but attains much smaller speaking speed error on validation set. This validates the effectiveness of joint training the speaking speed predictor and the whole model.

\section{Conclusion} \label{sec:conclusion}
In this work, we propose a novel non-AR end-to-end TTS model, VARA-TTS, generating mel-spectrogram from text with VDVAE and residual attention mechanism. 
The hierarchical latent variables from VDVAE are used as queries for the residual attention module.
The residual attention module is able to generate the textual-to-acoustic alignment in a layer-by-layer coarse-to-fine manner. 
Experimental results show that VARA-TTS attains better perceptual results than BVAE-TTS at a similar inference speed, and is 16x speed-up for inference over Tacotron 2 with slightly inferior performance in naturalness.
We also demonstrate its extensibility to a multi-speaker setting.

VARA-TTS should be easily extended to text-to-waveform by adding more layers. However, we find that it is hard to optimize in our preliminary experiments. The model is able to learn clear alignment between text and waveform, but it can not generate intelligible waveform. 
This is consistent with the results reported in \cite{child2020very} where VDVAE can not generate consistent and sharp $1024\times1024$ images.
We propose detailed $\operatorname{KL}$ gain for VDVAE, which can avoid non-informative hierarchical latent variables. It is interesting to analyze it theoretically and we leave this for the future work.

\bibliography{paper}
\bibliographystyle{icml2021}

\end{document}


\onecolumn

\icmltitle{Supplement Materials}

\vskip 0.3in




\appendix

\counterwithin{figure}{section}

\section{Auxiliary speaking speed predictor proof}
\label{supp:proof}
Minimizing the prediction error for speaking speed $\mathrm{d}$ from the top-most latent variables $\mathbf{z}_0$ paves the way to maximizing a lower bound of their mutual information.

\textbf{Proof}
\begin{align*}
    I(\mathbf{z}_0;\mathrm{d}) &= -\mathbb{H}(\mathrm{d}|\mathbf{z}_0) + \mathbb{H}(\mathrm{d}) \\
             &= \mathbb{E}_{\mathbf{z}_0 \sim p(\mathbf{z}_0)} \left[ \mathbb{E}_{\mathrm{d} \sim p(\mathrm{d}|\mathbf{z}_0)} \left[ \log p(\mathrm{d} | \mathbf{z}_0) \right] \right]  + \mathbb{H}(\mathrm{d}) \\
                 &= \mathbb{E}_{\mathbf{z}_0 \sim p(\mathbf{z}_0)} \left[ D_{\operatorname{KL}}(p(\mathrm{d}|\mathbf{z}_0) || q(\mathrm{d}|\mathbf{z}_0)) + \mathbb{E}_{\mathrm{d} \sim p(\mathrm{d}|\mathbf{z}_0)} \left[ \log q(\mathrm{d}| \mathbf{z}_0) \right] \right]  + \mathbb{H}(\mathrm{d}) \\
             &\geq \mathbb{E}_{\mathbf{z}_0 \sim p(\mathbf{z}_0)} \left[ \mathbb{E}_{\mathrm{d} \sim p(s|\mathbf{z}_0)} \left[ \log q(\mathrm{d}| \mathbf{z}_0) \right] \right] + \mathbb{H}(\mathrm{d}) \\
             &= \mathbb{E}_{\mathrm{d} \sim p(\mathrm{d})} \left[ \mathbb{E}_{\mathbf{z}_0 \sim p(\mathbf{z}_0|\mathrm{d})} \left[ \log q(\mathrm{d}| \mathbf{z}_0) \right] \right] + \mathbb{H}(\mathrm{d})
\end{align*}

$I(\mathbf{z}_0; \mathrm{d})$ is the mutual information between $\mathbf{z}_0$ and $\mathrm{d}$. $\mathbb{H}(\mathrm{d}|\mathbf{z}_0)$ is the conditional entropy of $\mathrm{d}$ given $\mathbf{z}_0$ and $\mathbb{H}(\mathrm{d})$ is the 
entropy of $\mathrm{d}$. Each piece of datum ($\mathbf{x}$ and $\mathbf{t}$) has its corresponding $\mathrm{d}$. $p(\mathbf{z}_0)$ is the prior distribution of $\mathbf{z}_0$
given $\mathbf{t}$, which omitted for conciseness. $p(\mathrm{d})$ is the prior of $\mathrm{d}$. It is not actually sampled. $\mathrm{d}$ is sampled when a piece of datum is sampled.
$q(\mathrm{d}|\mathbf{z}_0)$ is the output distribution of an auxiliary speaking speed predictor. $p(\mathbf{z}_0|\mathrm{d})$ is actually the 
posterior of $\mathbf{z}_0$ given $\mathbf{x}$ and $\mathbf{t}$.

\section{Alignments for different values of $\beta$ and $\gamma$.} \label{alignments}
\begin{figure*}[h]
        \centering
         \subfloat{%
              \includegraphics[height=12cm]{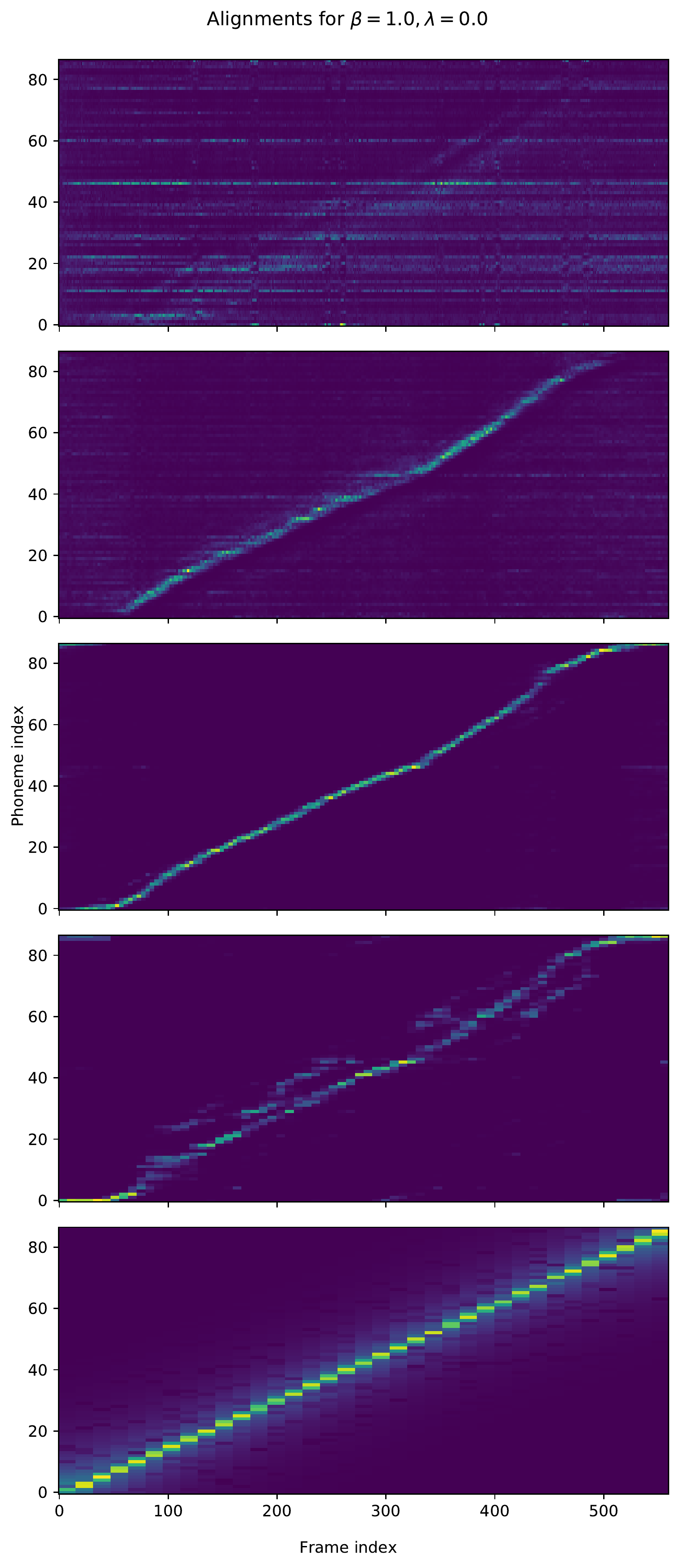}%
              \label{fig:left}%
         } 
           \quad
           \subfloat{%
              \includegraphics[height=12cm]{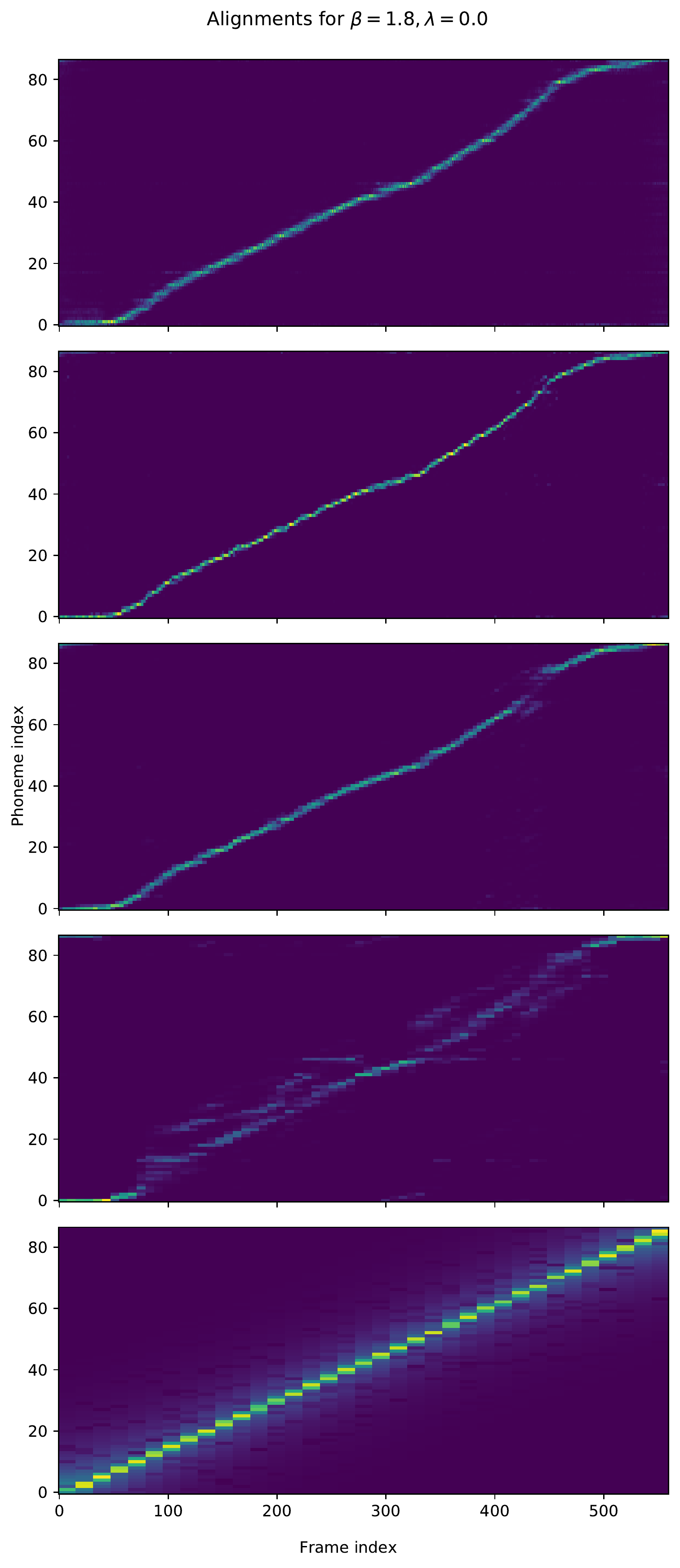}%
              \label{fig:middle}%
           }
           \quad
           \subfloat{%
              \includegraphics[height=12cm]{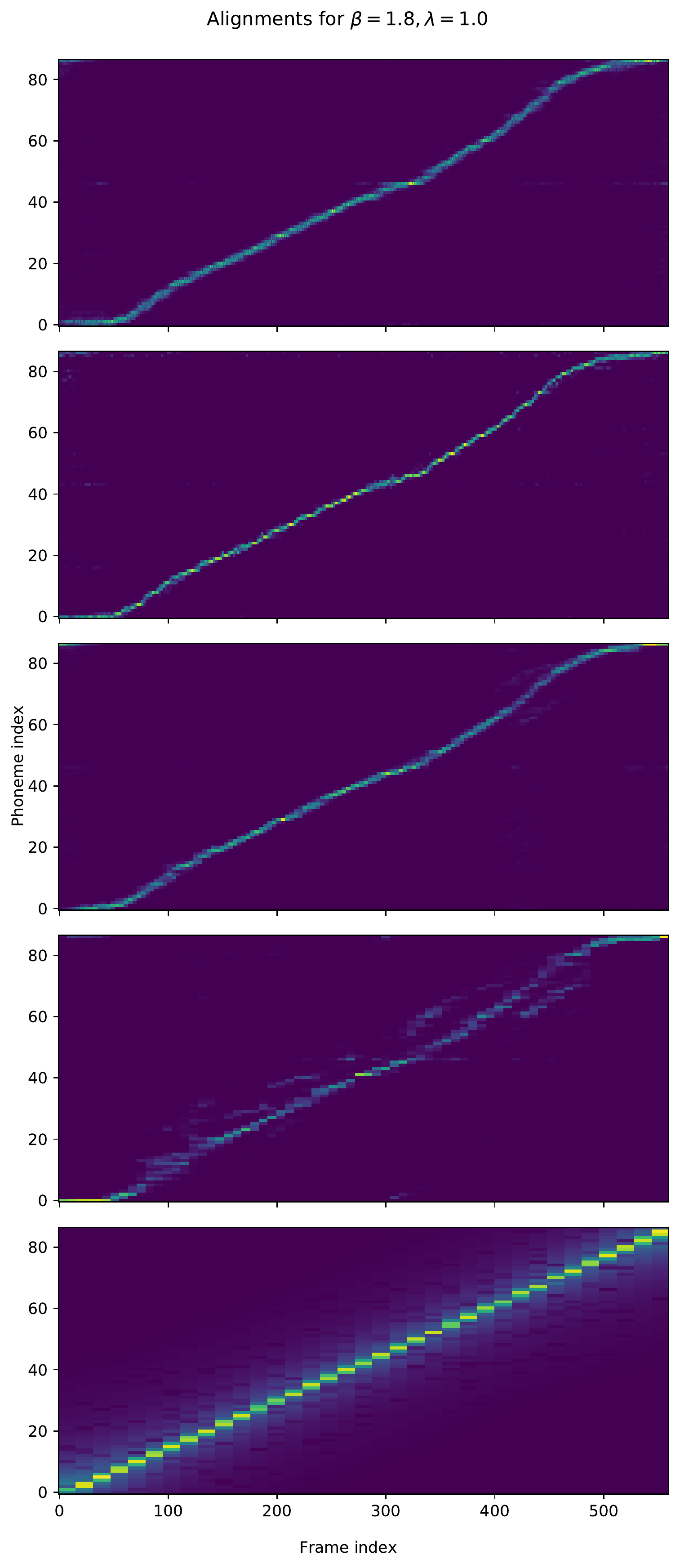}%
              \label{fig:right}%
           }
           	\vspace{-0.45cm}
           \caption{Alignments for different values of $\beta$ and $\lambda$. All the alignments are interpolated to the highest temporal resolution. 
           The figures from bottom row to top row correspond to coarse to fine hierarchical layers. 
           The alignments on the bottom row are diagonal alignments generated according to lengths of text and acoustic features.
           The following alignments are generated by residual attention mechanism. The alignments become clearer at finer hierarchical layers when $\beta=1.8$. However,
           when $\beta=1.0$, the alignments at finer hierarchical layers become blurry.
           When $\lambda=1.0$, the alignments at coarse layers is clearer than those when $\lambda=0$.}
           \label{fig:alignments}
    \end{figure*}

Figure \ref{fig:alignments} shows alignments for different values of $\beta$ and $\gamma$.

\section{Detailed model configuration}
Details of some key hyperparameters in VARA-TTS are listed in Table~\ref{table:hps_detail}. For the remaining model configuration, please refer to the source code accompanied with this manuscript.

\begin{table*}[ht]
\centering
\caption{Hyperparameters in VARA-TTS.}
\begin{tabular}{l|l|l}
\hline
\textbf{Hyperparameter name} & \textbf{VARA-TTS(EN)} & \textbf{VARA-TTS(ZH)} \\ \hline
Number of mel banks in mel spectrogram & 80 & 80 \\ \hline
Mel spectrogram pre-conv layer & Conv1D with k=11 and c=384 & Conv1D with k=11 and c=512 \\ \hline
Number of phoneme tokens & 55 & 148 \\ \hline
Text embedding dimension & 384 & 384 \\ \hline
Total number of bottom-up stacks & 6 & 7 \\ \hline
Number of res blocks in each bottom-up stack & 4/6/8/12/9/5 & 6/12/16/10/8/8/5 \\ \hline
Temporal reduction rate in each bottom-up stack & repeat/2/2/2/2/1 & repeat/2/2/2/2/2/1 \\ \hline
Bottom-up residual block conv dimensions & 384/96/96/384 & 512/128/128/512 \\ \hline
Number of heads in multihead attention module & 8 & 8 \\ \hline
Attention dimension & 384 & 384 \\ \hline
Latent variable dimension & 16 & 16 \\ \hline
Text encoder conv dimensions & 384/96/96/384 & 384/96/96/384 \\ \hline
Number of speakers & 1 & 7 \\ \hline
Speaker embedding dimension & - & 384 \\ \hline
\end{tabular}
\label{table:hps_detail}
\end{table*}

\section{Speaking speed predictor network structure}

\begin{figure}[t]
    \centering
	\includegraphics[height=4cm]{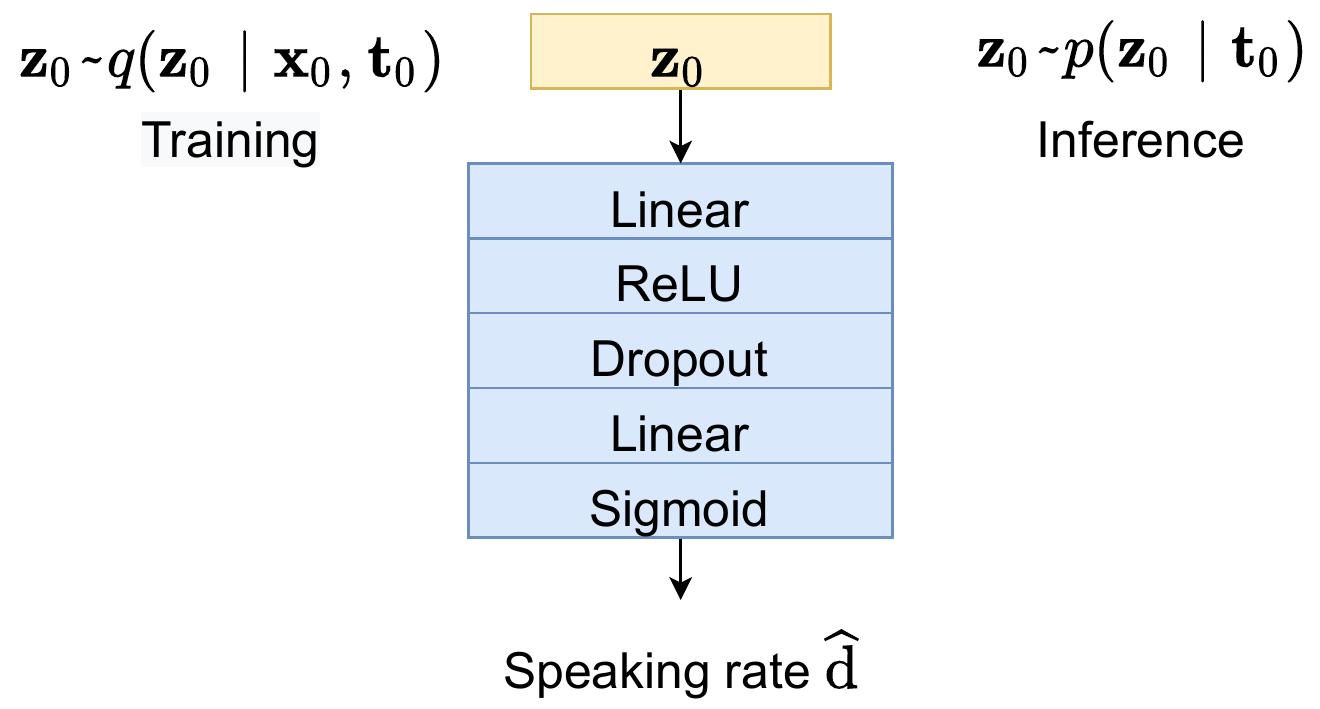}		
	\vspace{-0.45cm}
	\caption{Network detail of speaking speed predictor in VARA-TTS}
	\label{fig:d_predictor}
\end{figure}

Network structure of the speaking speed predictor used in VARA-TTS is illustrated in Figure \ref{fig:d_predictor}.

